\documentclass[12pt]{article}
\usepackage{times}
\usepackage{geometry}
\usepackage{graphicx}
\usepackage[utf8]{inputenc}
\usepackage{enumitem,amssymb}
\usepackage{ragged2e}
\newlist{thematic}{itemize}{8}
\setlist[thematic]{label=$\square$}
\usepackage{pifont}

\usepackage{amsmath,natbib,graphicx}
\begin{document}
\graphicspath{ {./} }
\DeclareGraphicsExtensions{.pdf,.eps,.png}
\raggedright
\large
\begin{center}
A Statistical Comparative Planetology Approach to Maximize the Scientific Return of Future Exoplanet Characterization Efforts \linebreak
\end{center}
\normalsize

\noindent \textbf{Thematic Area:}  Planetary Systems \hspace*{10pt}  \linebreak
  
\textbf{Principal Author:}

Name: Jade H. Checlair
 \linebreak						
Institution: University of Chicago
 \linebreak
Email: jadecheclair@uchicago.edu
 \linebreak
 
\textbf{Co-authors:} Dorian S. Abbot (University of Chicago), Robert J. Webber (New York University), Y. Katherina Feng (University of California, Santa Cruz), Jacob L. Bean (University of Chicago), Edward W. Schwieterman (University of California, Riverside), Christopher C. Stark (Space Telescope Science Institute), Tyler D. Robinson (Northern Arizona University), Eliza Kempton (University of Maryland)
\linebreak
 
\textbf{Co-signers:} Olivia D. N. Alcabes (University of Chicago), Daniel Apai (University of Arizona), Giada Arney (NASA Goddard), Nicolas Cowan (McGill University), Shawn Domagal-Goldman (NASA Goddard), Chuanfei Dong (Princeton University), David P. Fleming (University of Washington), Yuka Fujii (Tokyo Institute of Technology), R.J. Graham (University of Oxford), Scott D. Guzewich (NASA Goddard), Yasuhiro Hasegawa (Jet Propulsion Laboratory, California Institute of Technology), Benjamin P.C. Hayworth (Pennsylvania State University), Stephen R. Kane (University of California, Riverside), Edwin S. Kite (University of Chicago), Thaddeus D. Komacek (University of Chicago), Ravi K. Kopparapu (NASA Goddard), Megan Mansfield (University of Chicago), Nadejda Marounina (University of Chicago), Benjamin T. Montet (University of Chicago), Stephanie L. Olson (University of Chicago), Adiv Paradise (University of Toronto), Predrag Popovic (University of Chicago), Benjamin V. Rackham (University of Arizona), Ramses M. Ramirez (Earth-Life Science Institute), Gioia Rau (NASA Goddard), Chris Reinhard (Georgia Institute of Technology), Joe Renaud (George Mason University), Leslie Rogers (University of Chicago), Lucianne M. Walkowicz (The Adler Planetarium), Alexandra Warren (University of Chicago), Eric. T. Wolf (University of Colorado)
\linebreak

\begin{center}
\textit{White paper submitted in response to the solicitation of feedback for the “2020 Decadal Survey” by the National Academy of Sciences.} \linebreak
\end{center}

\pagebreak
\begin{center}
\textbf{Abstract:}
\justify
Provided that sufficient resources are deployed, we can look forward to an extraordinary future in which we will characterize potentially habitable planets. Until now, we have had to base interpretations of observations on habitability hypotheses that have remained untested. \textit{To test these theories observationally, we propose a statistical comparative planetology approach to questions of planetary habitability.} The key objective of this approach will be to make quick and cheap measurements of critical planetary characteristics on a large sample of exoplanets, exploiting statistical marginalization to answer broad habitability questions. This relaxes the requirement of obtaining multiple types of data for a given planet, as it allows us to test a given hypothesis from only one type of measurement using the power of an ensemble. This approach contrasts with a “systems science” approach, where a few planets would be extensively studied with many types of measurements. A systems science approach is associated with a number of difficulties which may limit overall scientific return, including: the limited spectral coverage and noise of instruments, the diversity of exoplanets, and the extensive list of potential false negatives and false positives. A statistical approach could also be complementary to a systems science framework by providing context to interpret extensive measurements on planets of particular interest. We strongly recommend future missions with a focus on exoplanet characterization, and with the capability to study large numbers of planets in a homogenous way, rather than exclusively small, intense studies directed at a small sample of planets.
\end{center}
\pagebreak

\section{The importance of a statistical approach in studying habitable exoplanets}
\justify						
\textbf{The systems science framework:} When studying habitable exoplanets, planetary scientists can be intuitively drawn to a “systems science” approach, which aims to reveal the various systems operating on individual planets using empirical data and theoretical modeling \citep[e.g.,][]{robinson2011earth,robinson2014detection}. A recent example of a “systems science” study is the Juno mission, where a single space probe was equipped with multiple instruments to obtain as many different types of data about Jupiter as possible. In this context, atmospheric biosignatures have been investigated, with the hope that they could prove that a planet is inhabited \citep[e.g.,][]{seager2005vegetation,meadows2008planetary,seager2010exoplanet}. With a systems science approach, only a few planets would be observed with multiple measurement techniques, making it well-suited for Solar System planets.\\	

\noindent \textbf{Limitations of the systems science framework:} Exoplanets present new challenges and opportunities that may make solely using a systems science approach less ideal. Terrestrial exoplanets are expected to be very diverse in the composition of their interiors and atmospheres. In particular, initial volatile inventory and planetary mass,  as well as the subsequent evolution of volatile delivery, is expected to result in a myriad of planetary interiors and atmospheres \citep{meadows2005modelling, bond2010compositional}. This will make it difficult to build an instrument that would allow us to obtain desirable measurements for a wide variety of them. Theoretical models are benchmarked on Solar System planets \citep[e.g.,][]{robinson2011earth,robinson2014detection}, but the great diversity of terrestrial exoplanets will present a challenge that may undermine their accuracy. Another difficulty is the inherent practical limitations of instruments. The spectral coverage will be limited, making certain biosignatures inaccessible, and the signal-to-noise ratio may not be sufficient to detect weaker signatures \citep{bolcar2017large,mennesson2016habitable}. A third major limitation of this approach is that there exist many false positive and false negative scenarios, some of which we have already identified and some of which remain unknown, that will make any biosignature difficult to interpret \citep[e.g.,][]{domagal2014abiotic,luger2015extreme,reinhard2017false}. \textit{Solely using a systems science approach therefore risks resulting in little convincing evidence of habitability.}\\
				 	 	 			
\noindent \textbf{A statistical approach:} In the search for life, the main advantage of exoplanets over solar system planets is their number \citep{cowan2015characterizing, NAP25187}. For example, while there is only one Jupiter, dozens of hot Jupiters have already been detected. To exploit this opportunity, \citet{bean2017statistical} proposed a statistical comparative planetology approach to questions of planetary habitability. The most well-known example of a statistical approach in astronomy is the Hertzsprung-Russell diagram, which allowed us to learn about the Sun by making low-precision measurements of a large sample of different stars \citep{hertzsprung1905stralung,russell1912relations}. Similarly, the great diversity of exoplanets is an opportunity for this approach, rather than a challenge, as it samples the naturally occurring phase space and allows us to statistically marginalize over the large uncertainties inherent to terrestrial planets. The large number of exoplanets will be leveraged by making relatively cheap and quick low-precision measurements on as many of them as possible to definitively test specific hypotheses related to planetary habitability. When we consider a planet of greater interest, the statistical information that will have been obtained will give context for the interpretation of measurements of that planet. Results from other observations can be also be incorporated afterwards to further increase the value of each survey, even if different objects are observed \citep[e.g.,][]{schwartz2015balancing}. \textit{The principal advantage of a statistical approach is that we will be able to obtain robust answers to specific habitability hypotheses, greatly increasing the return of future exoplanet characterization efforts.}
\begin{figure}[ht!]
\begin{center}
  \label{fig:StatsVSSystem}
  \includegraphics[width=0.4\textwidth]{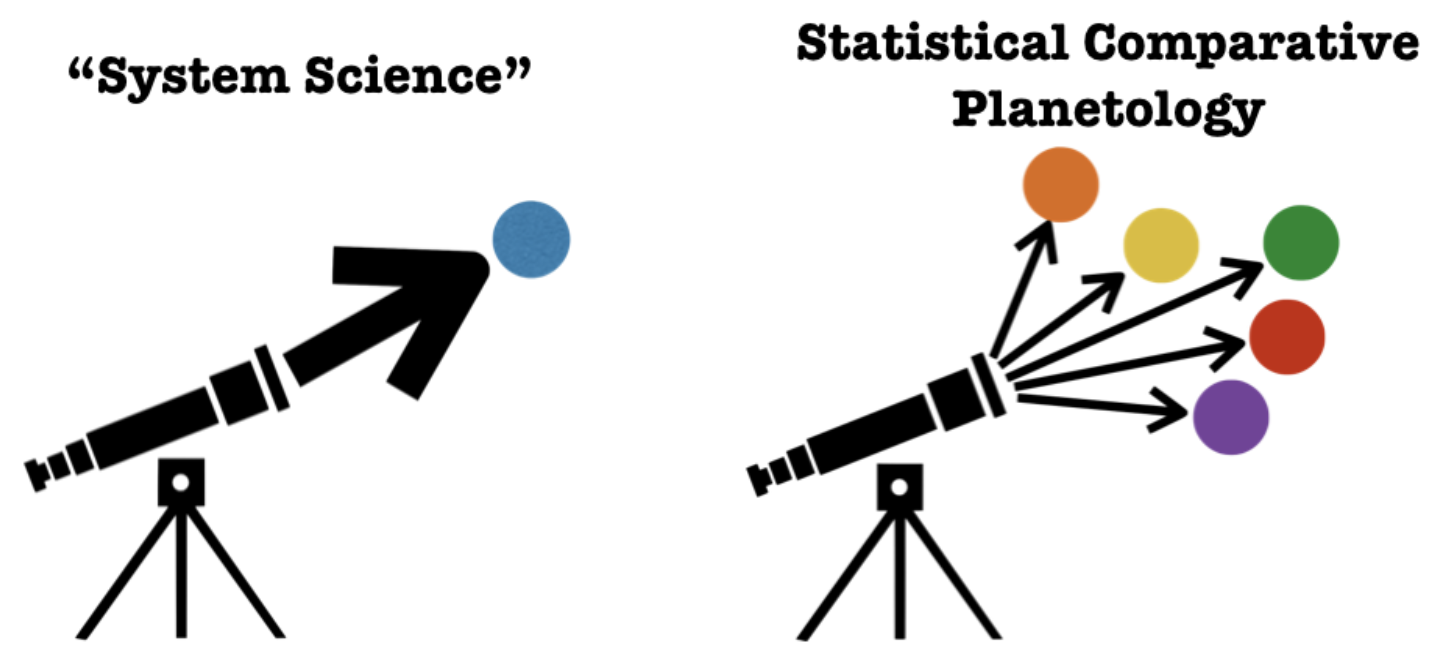}
    \caption{\textbf{This figure illustrates the difference between a system science approach, where a planet is characterized in great detail with many types of measurements, and a statistical approach, where each type of measurement is applied widely on a large sample of exoplanets.}}
  \end{center}
\end{figure}

\section{Case Study: Testing the habitable zone concept statistically}
\justify
The habitable zone concept is one of the main tools used to characterize potentially habitable exoplanets. Despite this, the habitable zone and its limits have not been tested observationally. A statistical approach will present us with the opportunity to test this concept, and further our understanding of the processes governing habitability. \\
\\
\noindent \textbf{Testing the limits of the habitable zone:} 

\indent Traditional habitable zone theory \citep{kasting1993habitable} assumes that terrestrial planets are able to maintain surface liquid water inside the habitable zone. If this assumption is correct, two testable predictions can be made. First, the abundance of water vapor should be greater inside than outside the habitable zone. This is because terrestrial planets inside the inner edge are expected to have lost all of their water vapor as a result of a runaway greenhouse process, while those orbiting outside the outer edge are expected to have all of their surface liquid water frozen. Second, terrestrial planets inside the habitable zone should tend to have a lower albedo than frozen planets outside the outer edge. If these predictions are correct, we should be able to detect threshold distances where the albedo and water vapor concentration increases using a large sample of planets.

Planetary color has been previously proposed as a method for discriminating Earth-like worlds from other planetary objects \citep{crow2011views, traub2003colors} and optimized photometric bands for identifying Earths in future space-based surveys have been calculated \citep{krissansen2016pale}. In a statistical comparative planetology frame, the carbonate-silicate cycle makes general predictions of planetary color within the inner and outer boundaries just as it does for albedo and the presence or absence of water vapor. At the inner boundary, planets should transition from bright and gray post-runaway atmospheres with volcanic condensates (e.g., Venus) to darker and bluer worlds with enhanced Rayleigh scattering from clear-sky paths to the surface (due to partial cloud cover from an active hydrological cycle). Additionally, the carbonate-silicate cycle predicts that CO$_2$ concentrations will rise with increasing distance from the host star and decreasing S$_{eff}$ until very high CO$_2$ levels are reached at the maximum greenhouse limit, which defines the outer edge \citep{kasting1993habitable, kopparapu2013habitable}. Habitable planets near the outer edge will be brighter than planets at the inner edge but they should also be bluer due to enhanced Rayleigh scattering from a larger atmospheric mass. In other words, planetary “blueness” should increase as a function of decreasing S$_{eff}$ from the inner to the outer boundaries. Frozen, ice-covered planets outside the outer boundary should be substantially less blue than planets just inside it because a collapse of CO$_2$ in the atmosphere would reduce atmospheric scattering (residual N$_2$ may remain, as N$_2$ has a low deposition temperature). Dry and barren planets like Mars should also be distinguishable from planetary color because most oxidized minerals have blue-absorption coupled with increasing spectral albedos into the red and infrared \citep{baldridge2009aster, clark2007usgs}. Planetary color is likely less expensive to secure than spectra with a space-based imaging survey and it may be possible to design optimized band passes for identifying the inner and outer edge transitions.  
\linebreak

\noindent \textbf{Testing climate regulation within the habitable zone:}

Traditional habitable zone theory \citep{kasting1993habitable} predicts that the surface temperature of habitable planets is regulated inside the habitable zone, allowing for surface liquid water. This theory assumes that the silicate-weathering feedback \citep{Walker-Hays-Kasting-1981:negative} regulates the atmospheric CO$_2$ of planets within the habitable zone (Figure 2) through a stabilizing negative feedback. As a planet’s surface temperature decreases, the weathering rate (intake of CO$_2$ by the crust) slows, allowing CO$_2$ to accumulate in the atmosphere and resulting in an increase in surface temperature. This feedback significantly extends the outer edge of the habitable zone, from 1.01 AU \citep{hart1979habitable} to 1.67 AU \citep{kasting1993habitable}, where outer planets build dense CO$_2$ atmospheres to maintain habitable surface conditions. It is also believed to be responsible for allowing Earth to escape Snowball Earth events. There is some non-definitive evidence that this feedback has worked in Earth’s history \citep[e.g.,][]{stolper2016pleistocene}, but it is untested in an exoplanet context. 
\begin{figure}[ht!]
\begin{center}
  \label{fig:kastingSW}
  \includegraphics[width=0.4\textwidth]{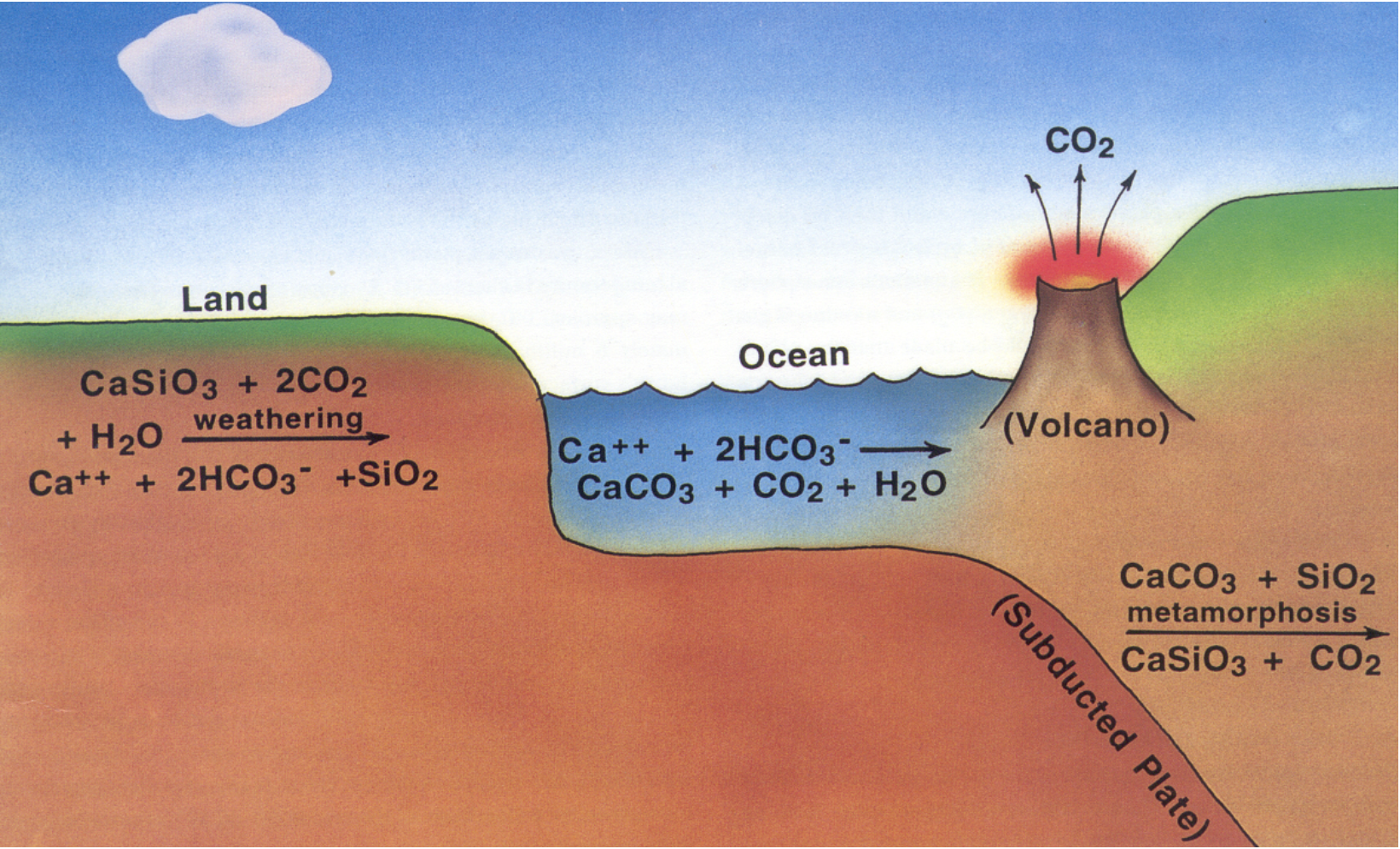}
    \caption{\textbf{The silicate-weathering feedback: Atmospheric CO$_2$ dissolves in rainwater, forming carbonic acid, which reacts with continental silicate rocks to form products that move into the oceans, and are then carried down subduction zones, where they release gaseous CO$_2$ which returns to the atmosphere through volcanoes. Kasting (1993).}}
  \end{center}
\end{figure}

\subsection{Worked Example: Testing the silicate-weathering feedback}
We propose a test for the operation of the silicate-weathering feedback using future instruments. To determine whether such a test is feasible, we calculate how many planets we would need to observe to conduct it. We do this by calculating the CO$_2$ that would be necessary to maintain habitable surface conditions as a function of stellar irradiation received by the planet, given uncertainty in planetary and atmospheric parameters as well as observational uncertainty. An idealized example of this calculation from \citet{bean2017statistical}  is shown in Figure 3. The salient point from this plot is that the mean CO$_2$ of planets decreases as the stellar irradiation they receive increases in order to maintain a roughly constant surface temperature, and that this trend could be observable if enough planets are measured to average over variation in planetary parameters. 

\begin{figure}[ht!]
\begin{center}
  \label{fig:bean2017}
  \includegraphics[width=0.4\textwidth]{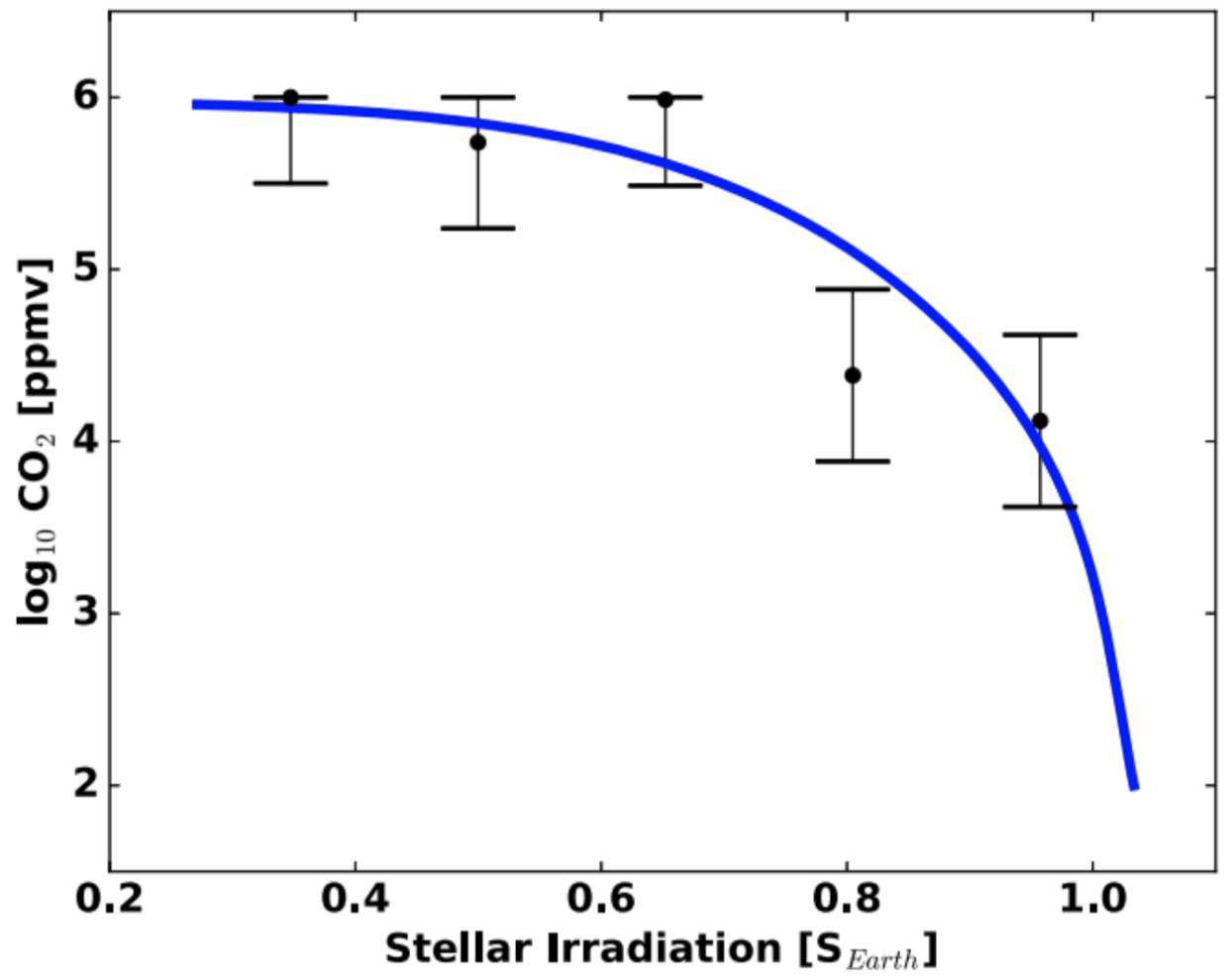}
    \caption{\textbf{This plot shows how the silicate-weathering feedback hypothesis, which assumes a decrease in atmospheric CO$_2$ as stellar irradiation increases, could be tested on exoplanets. The blue curve shows the predicted CO$_2$ needed to maintain a surface temperature of 290 K.}}
  \end{center}
\end{figure}

To determine how many planets are needed to carry out this test, we consider an optimistic and a pessimistic case for parameter variation. For our optimistic parameters, we choose means and standard deviations such that values typical of Earth-like planets in the habitable zone with a functioning silicate-weathering feedback are sampled \citep{krissansen2018constraining, olson2018earth, rogers2015most}. For our pessimistic assumptions, we increase our standard deviations such that the values sampled could represent any terrestrial planet in the habitable zone. For each parameter combination that we consider, we use the Clima 1D radiative-convective model \citep{kopparapu2013habitable} to compute vertical profiles of temperature and water vapor. We then feed these profiles into the Spectral Mapping Atmospheric Radiative Transfer (SMART) model \citep{meadows1996ground}, which can accurately calculate radiative fluxes, including the effects of clouds. We use SMART to calculate the outgoing longwave radiation (OLR) and planetary albedo ($\alpha$). From these we will infer the effective stellar flux at which planetary energy balance would be achieved, defined as S$_{eff}$ = 4 OLR / (1-$\alpha$). After generating S$_{eff}$ data, we built a low-order statistical model that predicts S$_{eff}$ based on the planetary parameters. Finally, to determine how many planets are needed to carry out this test, we use the Monte Carlo sampling technique. We simulate possible exoplanets by first drawing parameters from their distributions. We then calculate a value of S$_{eff}$ using our regression model for our parameter draw, including 0.5 decades of “instrumental” noise. We draw a fixed number of planets, calculate the resultant linear trend in CO$_2$ vs. S$_{eff}$, and calculate the p-value of the slope. We repeat this process 10$^5$ times and calculate the power of the test as the fraction of draws with a p-value less than 0.05, and then repeat for different numbers of planets. 

Our preliminary results for the optimistic case show that we would need to characterize 11 planets to ensure a power of 0.8, meaning an 80\% probability that if the feedback operates, we would be able to detect it if we measured 11 planets and the optimistic case is a realistic description of parameter variation (Figure 4). In our pessimistic case, we find that 51 planets would be needed to ensure a power of 0.8 (Figure 4). While a more accurate treatment of noise following \citet{feng2018characterizing} will be included in future work, these optimistic and pessimistic values provide us with lower and upper bounds on the sample size of potentially Earth-like planets needed to test for silicate-weathering feedback. 

These sample sizes are within reach. HabEx A, LUVOIR B, and LUVOIR A are estimated to detect ~8, ~30, and ~50 potentially Earth-like planets, respectively, within a 2 year blind survey (HabEx Interim Report, LUVOIR Interim Report, Stark private comm).  All of these missions have budgeted time in their nominal survey strategy to measure the parameters needed for the HZ limits study we describe, including orbits, colors, and at least cursory spectra to search for signs of water vapor. Additional spectral characterizations to measure the CO$_2$ abundances of these planets could enable the silicate-weathering feedback study we describe.

\begin{figure}[ht!]
\begin{center}
  \label{fig:powervsnp}
  \includegraphics[width=0.8\textwidth]{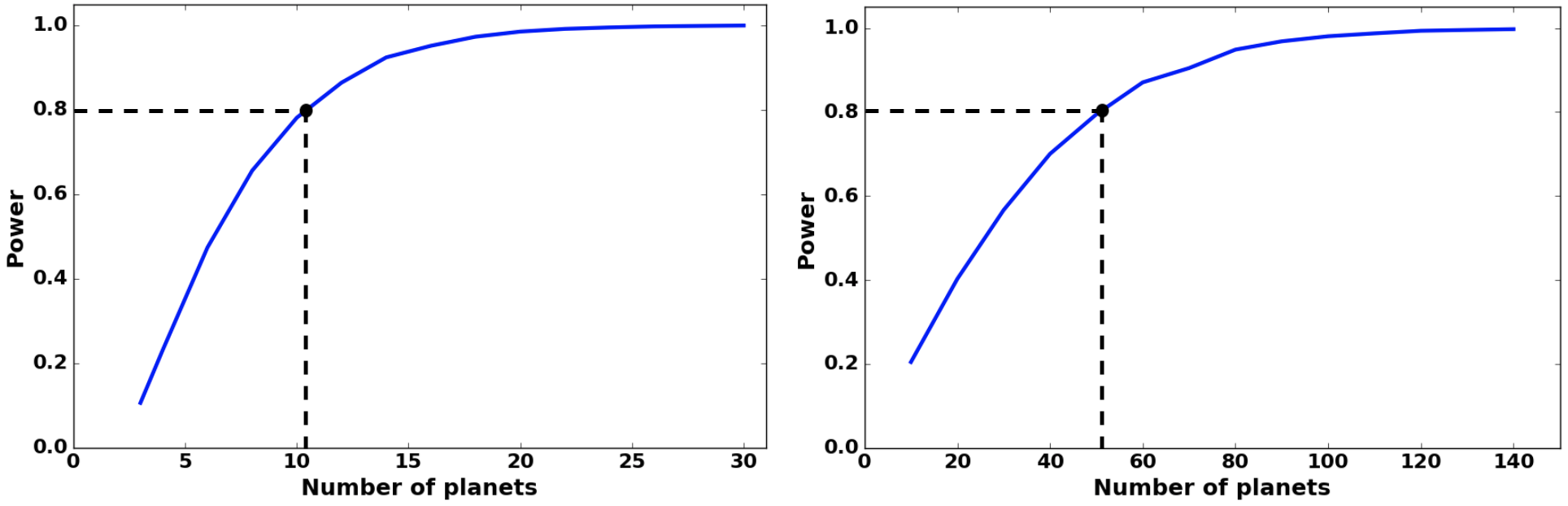}
    \caption{\textbf{The statistical power of a test to detect the silicate-weathering feedback as a function of the number of exoplanets observed, for an optimistic (left) and a pessimistic (right) case.}}
  \end{center}
\end{figure}

\section{Conclusions}

A statistical methodology promises to deliver definitive tests of fundamental habitability hypotheses, such as climate regulation within the habitable zone and the location of the limits of the habitable zone, if enough planets can be measured. The main advantage of this methodology is that it can confirm or falsify specific hypotheses concerning planetary habitability. In contrast, if biosignatures are searched for on only a few planets, there is a significant risk that the effort will end in negative or ambiguous results. Moreover, a statistical survey will provide context to interpret more detailed measurements of planets of particular interest. This work is of critical importance for maximal exploitation of limited observing time from future exoplanet characterization missions. \textit{Our recommendation for the decadal survey is that exoplanet characterization be a focus of future missions, and that these missions should be capable of studying a large number of habitable exoplanets to allow for statistically testing habitability hypotheses.}

\pagebreak

\bibliographystyle{ametsoc}
\bibliography{bib}

\end{document}